\documentclass[preprint,12pt]{elsarticle}

\usepackage{graphicx}
\usepackage{amsmath,amssymb}
\usepackage{amsthm}
\usepackage{arydshln}
\usepackage{rotate,rotating}
\usepackage{epsf,epic,eepic}
\usepackage{bmpsize}
\usepackage{multirow} 
\usepackage{bm}
\usepackage{hyperref}

\renewcommand{\vec}[1]{\mathbf{#1}}
\newcommand{\tens}[1]{\mathsf{#1}}
\newcommand{\tableheadseprule}{\noalign{\hrule height.375mm}}

\usepackage{amssymb}

\journal{Mathematical Biosciences}

\begin{document}

\begin{frontmatter}



\title{Coupled, multi-strain epidemic models of mutating pathogens}

 \author[AITHM]{Michael T. Meehan}
\author[Daniel]{Daniel G. Cocks}
\author[James]{James M. Trauer}
\author[AITHM]{Emma S. McBryde}

\address[AITHM]{Australian Institute of Tropical Health and Medicine, James Cook University, Townsville, Australia}
\address[Daniel]{College of Science and Engineering, James Cook University, Townsville, Australia}
\address[James]{School of Public Health and Preventive Medicine, Monash University, Melbourne, Australia}




\begin{abstract}
We introduce and analyze coupled, multi-strain epidemic models designed to simulate the emergence and dissemination of mutant (e.g. drug-resistant) pathogen strains. In particular, we investigate the mathematical and biological properties of a general class of multi-strain epidemic models in which the infectious compartments of each strain are coupled together in a general manner. We derive explicit expressions for the basic reproduction number of each strain and highlight their importance in regulating the system dynamics (e.g. the potential for an epidemic outbreak) and the existence of nonnegative endemic solutions. Importantly, we find that the basic reproduction number of each strain is independent of the mutation rates between the strains --- even under quite general assumptions for the form of the infectious compartment coupling. Moreover, we verify that the coupling term promotes strain coexistence (as an extension of the competitive exclusion principle) and demonstrate that the strain with the greatest reproductive capacity is not necessarily the most prevalent. Finally, we briefly discuss the implications of our results for public health policy and planning.

\end{abstract}


\end{frontmatter}






\section{Introduction}
\label{sec:Introduction}

The growing threat of antimicrobial drug resistance presents a significant challenge not only to the medical community, but the wider general population~\citep{mckenna2013antibiotic,laxminarayan2013antibiotic}: drug-resistant micro-organisms differ by the combination of antimicrobials to which they are resistant and susceptible (henceforward referred to as strains), which has important implications for their clinical and public health management. Indeed, resistant strains of infectious diseases are already endemic in many communities --- particularly in developing countries and lower socio-economic settings --- with new strains, that enjoy even more extensive resistance, continually emerging (see e.g.~\citep{world2015global}). 

Although resistance develops as a pathogen's natural biological response to antimicrobial treatment~\citep{walsh2000molecular,rodriguez2013antibiotics,blair2015molecular,walsh2015multiple}, the misuse and overuse of existing antimicrobials severely exacerbates the problem, rendering previously successful treatments ineffective. Consequently, we are now faced with a rapidly diminishing arsenal of effective therapies, with poor practice only accelerating us along this path~\citep{Neu1992}. As such, concern mounts that ``superbugs'' will emerge that are resistant to all available treatments, with many fearing that we are approaching the end of the antimicrobial era~\citep{hancock2007end}. 

To examine this growing public health concern, we introduce a general mathematical model designed to simulate the emergence and dissemination of mutant strains of infectious diseases. The formal framework assumes the form of a coupled multi-strain SIS/SIR/SIRS epidemic model~\citep{Anderson1992_infectious} in which individuals can transition between the various infectious compartments associated with each strain. This structure is intended to replicate the phenotypic phenomenon of \textit{amplification}, whereby individuals infected with a particular pathogen strain develop a new, mutant strain that is possibly resistant to some combination of antimicrobials. 

Several epidemiological investigations into the imminent threat of mutation and drug resistance, most often linked to a specific disease~\citep{hastings2000modelling,cohen2004modeling,blower2004modeling,thakur2013modelling},  and geographic setting~\citep{d2009modeling}, have already appeared in the literature (see also~\citep{LopezLozano200021,andersson2010antibiotic,zurWiesch2011236}).\footnote{Here we are focusing specifically on population-level models. For examples of within-host models see~\citep{day2004general,de2008multistrain,korobeinikov2014continuous}.} These papers~\citep{cohen2004modeling,blower2004modeling,boni2006epidemic}, which often utilize special cases of the general multi-strain network presented here, usually focus on specific epidemiological outcomes (e.g. interim disease burden, optimal intervention strategies~\citep{bonhoeffer1997evaluating}) whilst providing less mathematical detail on the dynamics of the model. Similarly, detailed mathematical analyses of multi-strain epidemic models~\citep{Ackleh2003,Bichara2013} primarily treat the infectious compartments as being parallel (i.e. uncoupled) and do not consider the possibility of infected patients ``amplifying'' to an alternate strain. The present paper attempts to bridge this gap by formally examining the mathematical implications of linking the various infectious compartments in a sufficiently general manner.\footnote{For earlier works on competitive exclusion, coexistence, and the co-evolution of hosts and parasites see~\citep{volterra1928variations,levin1970community,anderson1982coevolution,beck1984coevolution,Bremermann,andreasen1995pathogen,lipsitch1995host,lipsitch1996evolution,LIPSITCH199731}.}

Specifically, this article analytically examines the mathematical and biological aspects of the proposed coupled epidemic models. As such we focus our attention on the functional form of the basic reproduction number of each strain, discuss their importance in regulating the system dynamics and outline the necessary conditions for an epidemic outbreak. We also analyze the nature and structure of the asymptotic solutions of the system and explore how they relate to both the structure of the infectious compartment network and the relative magnitudes of the basic reproduction number of each strain. Numerical simulations of the model addressing disease-specific epidemiological issues will be the subject of future work.

To begin, we give a brief description of the coupled network of infectious compartments and their corresponding connectivity in sections~\ref{sec:coupling} and~\ref{sec:Inetwork}. Then, in sections~\ref{sec:model_description} and~\ref{sec:system_bounds}, we define our model parameters and introduce and analyze the set of differential equations governing the evolution of the SIS, SIR and SIRS systems. In section~\ref{sec:R0} we derive the set of basic reproduction numbers associated with each pathogen strain and demonstrate that these quantities represent threshold parameters that govern the dynamics of each system and, in particular, the potential for an epidemic outbreak (see sections~\ref{sec:epidemic_outbreak} and~\ref{sec:dynamics}).

We then determine the equilibrium states of each system in section~\ref{sec:equilibrium} and derive the necessary and sufficient conditions for the existence of nonnegative endemic solutions. In doing so we verify that the coupling term leads to the coexistence of several pathogen strains at the endemic equilibria and demonstrate that the strain with the greatest reproductive capacity is not necessarily the most prevalent.

Following this, in section~\ref{sec:un_comp}, we compare our findings with previous results derived within the context of uncoupled multi-strain epidemic models and examine how the coupling term modifies the system properties. Finally, in section~\ref{sec:conclusion}, we summarize our results, discuss their significance for public health policy and suggest directions for future work.

\section{Coupled infectious compartments}
\label{sec:coupling}

In this article we wish to analyze the dynamics of systems (i.e. populations) with several co-circulating pathogen strains. In particular, we are interested in the case where the co-circulating strains are related by genetic mutation. In this manner we are naturally led to coupled, multi-strain epidemic models (for examples of uncoupled multi-strain models see e.g.~\cite{Ackleh2003,Bichara2013}). 

To provide a rigorous framework, we will begin with a simple class of epidemic models in which the population is stratified into three broad categories: susceptible individuals, S; individuals who are both infected and infectious, I$_i$, where the subscript $i \in [1,n]$ labels infection with a particular pathogen strain and $n$ is the total number of possible strains; and individuals who have recovered from infection with strain $i$, R$_i$. 

For generality, we will consider an overarching structure that incorporates the standard SIS, SIR and SIRS epidemic models. These three models can be differentiated by their immunity-status post-infection. In the SIS model, infected individuals immediately return to the susceptible class upon recovery from infection and the R$_i$ category is redundant. In the other models, recovered individuals, R$_i$, enjoy immunity, which is permanent (SIR), or temporary (SIRS), against subsequent infection with any strain. Respectively, these models may, for example, be applicable to chlamydia (SIS), influenza and rubella (SIR), or tuberculosis (SIRS). Importantly, we assume that only individuals in S are susceptible to infection, and each strain $i$ dips from the same susceptible pool, S. 

\begin{figure}[htbp]
\centering
\includegraphics[width=0.6\textwidth]{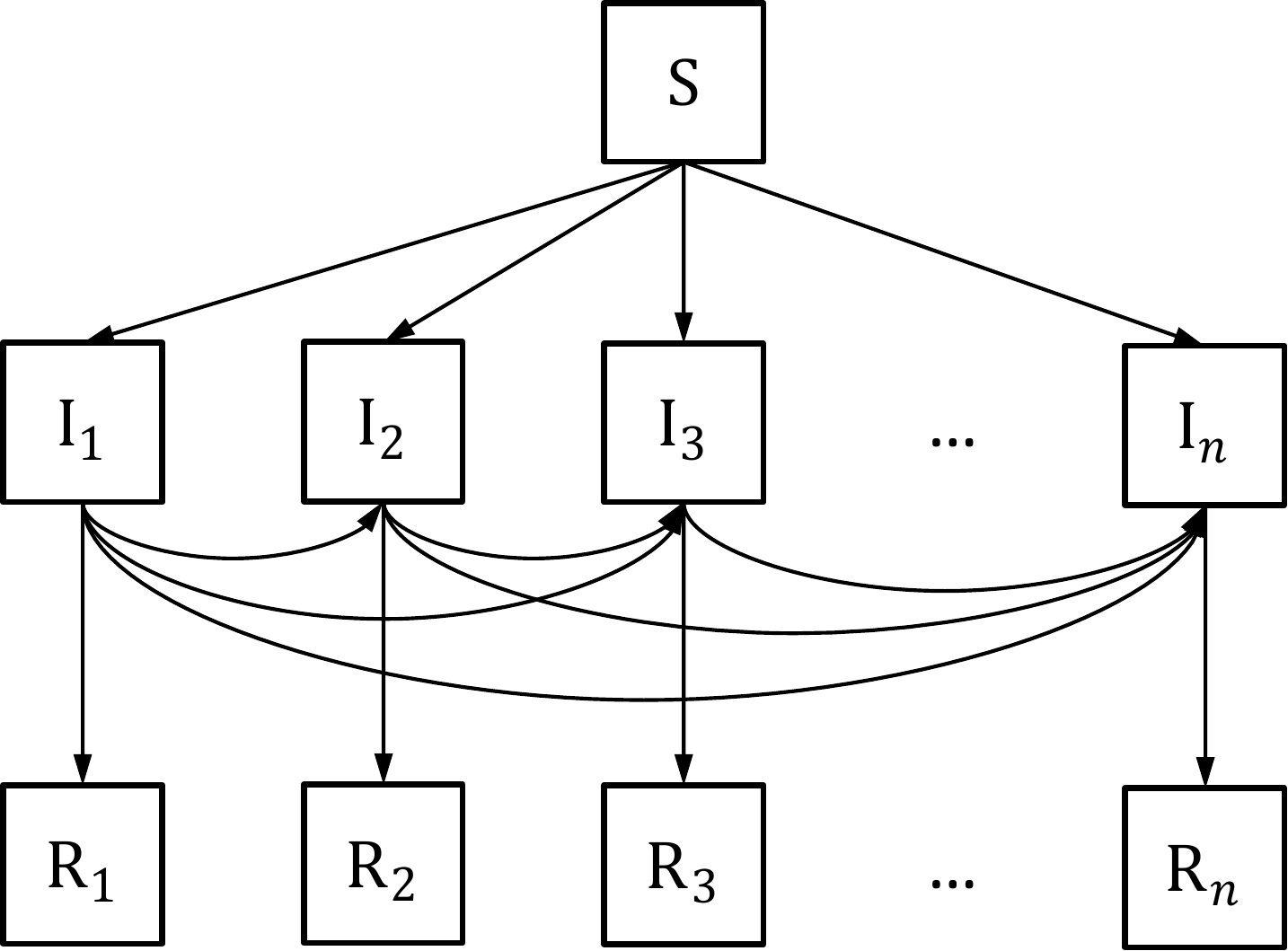}
\caption{Flow diagram for the SIR coupled multi-strain model.}
\label{fig:SIR_network}
\end{figure}

An example of the compartmental flow network of our multi-strain model is illustrated in figure~\ref{fig:SIR_network}.

The primary ingredient in our model is a coupling term between the various nodes of the infectious compartment network (see figure~\ref{fig:SIR_network}) which embodies the genetic relationship among the co-circulating strains. This coupling allows for individuals to flow directly between the infectious compartments without first returning to the susceptible (S) or recovered (R$_i$) classes --- a feature designed to replicate the \textit{in vivo} mutation of pathogens.

A fundamental assumption that we make at the outset is that genetic evolution is a \textit{time-forward process}.\footnote{Whilst it is possible to envision that a series of mutation events might lead to a particular genetic descendant reverting back to an earlier copy of one if its ancestors, such as a deletion followed by a restoring insertion, such \textit{reversions} are exceedingly unlikely and can be safely neglected.} In this manner we need only consider a unidirectional coupling between the nodes in the infectious compartment network (see figure~\ref{fig:SIR_network}). In effect, individuals can only transition from their current infected compartment to one that is ``downstream'', i.e. an individual currently in I$_i$ can only transition to the infected state I$_j$ if $j > i$. In the language of graph theory we would say that the I$_i$ compartments (in isolation from the S and R$_i$ compartments) represent a directed acyclic graph.\footnote{If all of the I$_j$ are directly accessible from I$_i$ whenever $j > i$ then the network is said to be \textit{complete}.} We emphasize that in this case the ordering of the I$_i$ compartments and their numbered labelling is important.

We stress however, that increasing strain number \textit{does not} necessarily imply increasing levels of drug resistance: although we expect that mutation events that confer some level of drug resistance will be preferentially selected under the applied pressure of antimicrobial treatment, such that there is a natural progression towards higher levels of resistance, we do not assume \textit{a priori} that mutant, descendant strains are more drug-resistant than their ancestors. To the contrary, a particular genetic descendant strain may become more drug-susceptible than its ancestor via deleterious hitchhiker mutations.

Another simple example in which the ordered infectious compartments do not correspond directly to increasing levels of drug resistance could be a construction that simply incorporates differences in clinical outcomes for a single genetic variant. Here we might consider several subsets of ``phenotypically'' identical strains (i.e. strains with identical pathogenic properties and responses to treatment) within the multi-strain network that only differ by their treatment \textit{rates}. The coupling term between the strains could then reflect the rate at which infected individuals are properly diagnosed and treated.

That said, the prime motivation behind our model structure is to simulate the recent outbreak of drug-resistant pathogen strains. In this case, individuals infected with a particular pathogen strain that mutates and acquires drug resistance, either through natural mutation or as a result of ineffective treatment, transition from their current infected state to one of the ``downstream'' infected compartments. This process is known as ``amplification'' (not to be confused with the notion of super-infection\footnote{Super-infection occurs when an individual infected with a particular strain (say, strain $j$) transmits the infection to another individual already infected with an alternate strain (strain $i \neq j$). This process, which is traditionally modelled by an additional bilinear interaction term of the form $\kappa_{ij} I_i I_j$, is not considered in the present framework.}). The ordering of the infected compartments in this case, and the unidirectional flow between them, then reflects (but is not strictly connected to) the level of drug resistance of the pathogen. 

For instance, we could construct a coarse model that groups strains solely by resistance profile, so that, for example, an infectious pathogen which has only $t$ antimicrobial treatments available can be modelled by a coupled $2^t$-strain network, with each strain corresponding to each of the $2^t$ possible antimicrobial resistance combinations. In this class of models we would consider tiers of drug resistance where the $m$-th tier contains $t\choose m$ strains that are resistant to $m$ different antimicrobial treatments.

In general, each of the scenarios described above can be constructed within the present framework simply by limiting the possible transitions within the infectious compartment network and selecting appropriate values of the epidemiological parameters.

\section{Connectivity of the infectious compartment network}
\label{sec:Inetwork}

In this section we formalize several definitions that will be used throughout the text and derive important relationships pertaining to the connectivity of the network of I$_i$ compartments (figure~\ref{fig:SIR_network}). These include the \textit{reachability} of compartment I$_j$ from compartment I$_i$ and the definition of \textit{ancestor} and \textit{descendant} strains. Primarily, this discussion is designed to provide the reader a deeper appreciation of the structure of the next-generation matrix and endemic equilibrium solutions (derived in sections~\ref{sec:R0} and~\ref{sec:equilibrium} respectively) of coupled multi-strain models. For a more detailed introduction to the basic definitions and concepts of graph theory see e.g.~\citep{biggs1993algebraic}.

To begin, we introduce the \textit{adjacency matrix} $\tens{G}$,
\begin{equation}
\tens{G} = 
  \begin{bmatrix}
    0        & 0  & 0    & \cdots & \hfill & 0 \\
    g_{21} & 0        & 0     & \hfill & \hfill  & \vdots \\
    g_{31} & g_{32} & 0         & \ddots & \hfill  & \hfill \\
		g_{41} & g_{42} & g_{43}  & 0      & \hfill & \hfill \\
		\vdots& \vdots & \hfill & \ddots & \ddots & \hfill\\
		g_{n1} & g_{n2} & \cdots & \hfill &  g_{n,n-1} & 0
  \end{bmatrix},\label{eq:matG}
\end{equation}
where each element $0 \leq g_{ji} \leq 1$ gives the proportion of individuals infected with strain $i$ that acquire strain $j$ through ineffective treatment, or the proportion of \textit{amplifying} individuals among those treated. Accordingly, we require $\sum_{j = 1}^n g_{ji} \leq 1$. In this way $g_{ji}$ is representative of the transition probability between compartments I$_i$ and I$_j$. (The sum of the transition proportions is less than or equal to one because not all treated patients are expected to amplify.) Furthermore, the strictly lower triangular structure of the adjacency matrix $\tens{G}$ reflects the unidirectional flow pattern of the I$_i$ network. 

To be careful, we point out that, alone, the quantities $g_{ji}$ do not represent transition probabilities themselves; this is because the amplification pathways only represent a subset of the possible transitions out of the infectious compartments in the complete SIS/SIR/SIRS model (see next section). It is only when we augment the $g_{ji}$ terms with the appropriate rate parameters describing the relative flows out of the infectious compartments (sections~\ref{sec:model_description} and~\ref{sec:R0}) that we obtain the probability of transitioning from infectious compartment I$_i$ to compartment I$_j$. However, for our present purposes, we assume in this section that the elements $g_{ji}$ \textit{can} be interpreted as transition probabilities and reserve the inclusion of the appropriate rate parameters for later sections. It will become clear that the present discussion remains valid despite the absence of the relevant rate parameters.


Assuming (for now) the elements $g_{ji}$ represent the probability of transitioning from compartment I$_i$ to compartment I$_j$ in a single `step', the probability of ever reaching compartment I$_j$ from compartment I$_i$ is given by the $(j,i)$-th element of the \textit{connectivity matrix} $\tens{P}_\infty$,\footnote{If the probability of transitioning from compartment I$_i$ to compartment I$_j$ in a single `step' is given by the $(j,i)$-th element of the transition matrix $\tens{G}$, then the probability of reaching compartment I$_j$ from compartment I$_i$ after exactly $k$ steps is simply equal to the $(j,i)$-th element of $\tens{G}^k$. Since $\tens{G}$ is nonnegative (i.e. $\tens{G} \geq 0$) we can generalize this result to determine that the probability of reaching I$_j$ in at most $m$ steps is given by
\begin{equation}
\tens{P}_m = \sum_{k = 0}^m \tens{G}^k.\nonumber
\end{equation}
Similarly, the interpretation of the $(j,i)$-th element of $\tens{P}_m$ is analogous to that of $\tens{G}^k$. $\tens{P}_\infty$ is found by taking the limit as $m\rightarrow \infty$ (which is guaranteed to exist since $\tens{G}$ is strictly lower triangular).}
\begin{equation}
\tens{P}_\infty = \sum_{k = 0}^\infty \tens{G}^k = \left(\tens{E} - \tens{G}\right)^{-1},\label{eq:con_mat}
\end{equation}
where $\tens{E}$ is the identity matrix. It follows then that compartment I$_j$ is \textit{reachable} from compartment I$_i$ if and only if the $(j,i)$ element of $\tens{P}_\infty$ is positive. Naturally, the connectivity matrix $\tens{P}_\infty$, which in this case is the inverse of a lower-triangular matrix, is itself lower-triangular.\footnote{This can be seen immediately by looking at the expansion 
\begin{equation}
\tens{P}_\infty = \left(\tens{E} - \tens{G}\right)^{-1}=\sum_{k = 0}^n \tens{G}^k = \tens{E} + \tens{G} + \tens{G}^2 +\ldots,\nonumber
\end{equation}
where each term in the series is lower triangular.} Moreover, the diagonal elements of $\tens{P}_\infty$, which correspond to the probability of reaching the infectious compartment an individual already occupies, are equal to one, as expected.

Continuing, we can use the definition of reachability to define the notion of \textit{descendant} and \textit{ancestor} strains. Specifically, strain $j$ is said to be a \textit{descendant} of strain $i$ if and only if infectious compartment I$_j$ is reachable from infectious compartment I$_i$ \textit{and} $j > i$. Similarly, if strain $j$ is reachable from strain $i$ \textit{and} $i < j$ then we say strain $i$ is an \textit{ancestor} of strain $j$. Note that a strain cannot be an ancestor or descendant of itself.

As a useful shorthand, we introduce the notation $j \triangleright i$ to denote that $j$ is a descendant of strain $i$ and $i \triangleleft j$ to denote that $i$ is an ancestor of strain $j$.\footnote{The direction of the triangles has been chosen to reflect the unidirectional flow pattern of the infectious compartment network and their numbered labelling. For example, strain $j$ can only be a descendant of strain $i$ if $j > i$.} More formally, we then have
\begin{equation}
j\triangleright i \iff i \triangleleft j \iff \left[\tens{P}_\infty - \tens{E}\right]_{ji} \neq 0.\nonumber
\end{equation}
Additionally, if $l$ is an ancestor of $i$ which itself is an ancestor of $j$, then $l$ is also an ancestor of $j$, i.e. $l \triangleleft i \triangleleft j \Rightarrow l \triangleleft j$. Of course the reverse is also true: $j \triangleright i \triangleright l \Rightarrow j\triangleright l$. Since the I$_i$ network is a directed acyclic graph, i.e. we only consider unidirectional flow, the concept of ancestors and descendants is self-consistent.

We define the quantity $d_i$ as the number of descendants that strain $i$ has, which is equal to the number of non-zero elements in the $i$-th column of $\tens{P}_\infty - \tens{E}$. 
Similarly, 
$a_i$ is defined to be the number of ancestors that strain $i$ has, which is equal to the number of non-zero elements along the $i$-th row of $\tens{P}_\infty - \tens{E}$.

As alluded to in the opening to this section, the connectivity matrix $\tens{P}_\infty$ and its underlying structure play a pivotal role in the nature and form of the model solutions. As we shall see, $\tens{P}_\infty$, or terms analogous to it, repeatedly appear in the coming sections when we calculate the next-generation matrix $\tens{K}$ (see section~\ref{sec:R0}) and the endemic equilibrium solutions (see section~\ref{sec:equilibrium}). To highlight this similarity we write the elements of $\tens{P}_\infty$ in terms of the generating function\footnote{To derive~\eqref{eq:pinfgen} we begin with the definition of the connectivity matrix $\tens{P}_\infty$ which we know must satisfy $\left(\tens{E} - \tens{G}\right)\tens{P}_\infty = \tens{E}$. Writing this equation out in component form we have
\begin{equation}
\left[\tens{P}_\infty\right]_{ji} - \sum_{k = i}^n g_{jk}\left[\tens{P}_\infty\right]_{ki} = \delta_{ji},\nonumber
\end{equation}
where $\delta_{ji}$ is the Kronecker-delta symbol which is equal to one if $i = j$ and zero otherwise, and we have used the fact that $\left[\tens{P}_\infty\right]_{ki} = 0$ for all $k < i$ (i.e. $\tens{P}_\infty$ is lower-triangular). For $j > i$ we recover~\eqref{eq:pinfgen}.}
\begin{equation}
[\tens{P}_\infty]_{ji} = \sum_{k = i}^n g_{jk} [\tens{P}_\infty]_{ki},\label{eq:pinfgen}
\end{equation}
where the initial value $[\tens{P}_{\infty}]_{ii} = 1$ (as mentioned above, $\left[\tens{P}_\infty\right]_{ji} = 0$ for all $j < i$). 


In section~\ref{sec:equilibrium} we will show that the endemic equilibrium solutions can be written in a similar form to~\eqref{eq:pinfgen} with the mutation rates $g_{jk}$ augmented by the appropriate rate parameters.

Naturally, if we take the limit $\tens{G} \rightarrow \tens{0}$ we remove all of the connections between the nodes of the infectious compartment network and the coupled multi-strain model reduces to the canonical uncoupled case. For this reason, the adjacency matrix, $\tens{G}$, plays a fundamental role in the analysis that follows.

\section{Model description and system equations}
\label{sec:model_description}

Returning to our description of the structure of the SIS, SIR and SIRS epidemic models, we adopt the convention that italicized letters denote the number of individuals in each of the corresponding categories so that, for example, $I_i$ represents the number of individuals in compartment I$_i$. 
Following this convention, the total population, $N(t)$, is given by\footnote{Note that $R_i(t) = 0$ for all $i$ in the SIS model.}
\begin{equation}
N(t) = S(t) + \sum_{i = 1}^n\left(I_i(t) + R_i(t)\right).\label{eq:Nsum}
\end{equation}

Demographically, we assume that individuals are recruited directly into the susceptible class, S, at a constant rate $\lambda$ and that they are removed from all classes through natural death at a per capita rate $\mu$. Furthermore, we assume that infected individuals, I$_i$, suffer a strain-specific mortality rate, $\phi_i$, in addition to the natural mortality rate $\mu$.\footnote{In the absence of infection-induced mortality ($\phi_i \rightarrow 0$) the natural mortality rate can be directly related to the average life expectancy $\ell$ via $\mu = 1/\ell$.} These assumptions allow us to immediately write down the differential equation for the total population, $N(t)$:
\begin{equation}
\frac{dN}{dt} = \lambda - \mu N - \sum_{i = 1}^n \phi_i I_i.\label{eq:Ndot}
\end{equation}

As for the time evolution of the S, I$_i$ and R$_i$ compartments themselves, we model the transmission of each pathogen strain assuming a bilinear form for the incidence of infection: $\beta_i S I_i$, where $\beta_i$ is a strain-specific transmission parameter. Upon infection with strain $i$, individuals immediately progress to the corresponding infectious compartment I$_i$. From here, there are several possible progression pathways. Firstly, individuals can either recover naturally, which occurs at a strain-specific per capita rate $\sigma_i$, returning to either the susceptible (S) or recovered (R$_i$) class, depending on whether the model is SIS or SIR/SIRS respectively. Secondly, individuals may receive treatment at the per capita rate $\alpha_i$: if treatment is unsuccessful, \textit{and} the patient acquires a mutant strain as a result of ineffective treatment, then a proportion $g_{ji}$ will progress to each of the downstream infectious compartments I$_j$ with $j > i$ (see section~\ref{sec:coupling}); the remaining proportion $\left(1 - \sum_{j = i+1}^n g_{ji}\right)$ for which treatment \textit{is} successful will progress to the susceptible (SIS model) or recovered (SIR/SIRS models) compartments.

\begin{table}
\caption{Model parameters}
\label{tab:model_parameters}       
\centering
\begin{tabular}{@{}lclr@{}}
\hline\noalign{\smallskip}
Type & Parameter  & Definition & Units \\[2pt]
\tableheadseprule\noalign{\smallskip}
\multirow{7}{*}{Primary rate} & $\lambda$ & Recruitment rate &  $PT^{-1}$ \\
 & $\mu$       & Mortality rate (per capita) &  $T^{-1}$ \\
 & $\beta_i$   & Transmission coefficient (per capita) &  $P^{-1}T^{-1}$ \\
 & $\alpha_i$ & Treatment rate (per capita) &  $T^{-1}$ \\
 & $\sigma_i$ & Natural cure rate (per capita) &  $T^{-1}$ \\
 & $\phi_i$     & Disease-induced mortality rate (per capita) &  $T^{-1}$ \\
 & $\gamma_i$ & Waning immunity rate (per capita) &  $T^{-1}$ \\
\\
 \multirow{2}{*}{Aggregate rate} & $\eta_i$     & Infected recovery rate (per capita) &  $T^{-1}$ \\
 & $\delta_i$  & Infected removal rate (per capita) &  $T^{-1}$ \\
\\
Proportion & $g_{ij}$        & Amplification proportion     & $-\quad$ \\
\\
Boolean & $\rho$       & Switch between SIS$\leftrightarrow$SIR(S) & $-\quad$\\ 
\\[1pt]\hline
\end{tabular}
\end{table}

Combining each of these assumptions with our parameter definitions, we find that our system is governed by the following set of differential equations:
\begin{align}
\frac{dS}{dt} &= \lambda - \mu S - \sum_{i = 1}^n \beta_i S I_i + \rho \sum_{i = 1}^n \eta_i I_i + \sum_{i = 1}^n \gamma_i R_i,\label{eq:dS}\\
\frac{dI_i}{dt} &= \beta_i S I_i - \delta_i I_i + \sum_{j = 1}^{i - 1} g_{ij} \alpha_j I_j,\label{eq:dIi}\\
\frac{dR_i}{dt} &= \left(1 - \rho\right)\eta_i I_i - (\mu + \gamma_i)R_i,\label{eq:dRi}
\end{align}
where $\eta_i$ and $\delta_i$ are aggregate rates given by
\begin{equation}
\eta_i = \sigma_i + \left(1 - \sum_{j = i + 1}^n g_{ji}\right)\alpha_i\quad \mbox{and} \quad \delta_i = \mu + \phi_i + \sigma_i + \alpha_i,
\end{equation}
and represent the flow of recovering individuals and outgoing infected individuals, respectively. The variable $\rho$ in~\eqref{eq:dS} and~\eqref{eq:dRi} is a boolean parameter used to switch between the SIS ($\rho = 1$) and SIR(S) ($\rho = 0$) models.\footnote{It is straightforward to generalize to the case where $\rho$ is strain-specific, such that particular strains may or may not have permanent or temporary immunity.} Similarly, by setting $\gamma_i = 0$ for all $i$ we obtain an SIS ($\rho = 1$) or SIR ($\rho = 0$) model. Importantly, we note that the equation for $dI_i/dt$ is invariant between the SIS, SIR and SIRS cases. As we shall see, this ensures that many of our findings pertaining to the functional form of the basic reproduction numbers of each strain, as well as the existence and direction of the asymptotic solution vectors, are generic amongst the models. Furthermore, in the limit $g_{ji}\rightarrow 0$ the system~\eqref{eq:dS}-\eqref{eq:dRi} reduces to an uncoupled multi-strain model. 


The system~\eqref{eq:dS}-\eqref{eq:dRi} can be written more succinctly in vector form as
\begin{align}
\frac{dS}{dt} &= \lambda - \mu S - S\vec{\beta}^\intercal \vec{I} + \rho\vec{\eta}^\intercal \vec{I} + \vec{\gamma}^\intercal\vec{R},\label{eq:dSmat}\\
\frac{d\vec{I}}{dt} &= \left(S\tens{B} - \tens{\Delta} + \tens{G}\tens{A}\right)\vec{I},\label{eq:dImat}\\
\frac{d\vec{R}}{dt} &= (1 - \rho)\tens{H}\vec{I} - (\mu\tens{E} + \tens{\Gamma})\vec{R},\label{eq:dRmat}
\end{align} 
where 
\begin{equation}
\vec{I} = [I_1, I_2,\ldots, I_n]^\intercal,\qquad \vec{R}=[R_1,R_2,\ldots,R_n]^\intercal,
\end{equation}
and we have introduced capital Greek letters to denote the diagonal matrices associated with each of the parameter vectors
\begin{align}
\tens{A} &= \mathrm{diag}(\vec{\alpha}),\quad \tens{B} = \mathrm{diag}(\vec{\beta}),\quad \tens{\Gamma} = \mathrm{diag}(\vec{\gamma}),\nonumber\\
\tens{\Delta} &= \mathrm{diag}(\vec{\delta}), \quad \tens{H} = \mathrm{diag}(\vec{\eta}).\label{eq:mat_def}
\end{align}
The final matrix, $\tens{G}$, is the lower triangular matrix containing the amplification proportions between the various infectious compartments, $g_{ji}$, defined by~\eqref{eq:matG}.

Notice that the coupling term $\tens{G}$ enters the differential system of equations through the product $\tens{G}\tens{A}$. Therefore, we have implicitly assumed that mutation events are directly related to the treatment rates $\alpha_i$. We stress however, that this assumption does not affect the general conclusions reached below, particularly those regarding the functional dependence of the basic reproduction numbers and the form of the endemic solutions. Even if we included an additional coupling term to represent natural mutation events, e.g. a term of the form $h_{ji}\sigma_i$ (analogous $g_{ji}\alpha_i$), our general conclusions would still hold (though the magnitudes of the endemic solutions would vary) provided this additional coupling was still unidirectional, i.e. that it represented a \textit{time-forward} evolutionary process. Therefore, we choose to stick with a single coupling term to avoid over-complicating the forthcoming analysis, keeping in mind that the results can be readily applied to the case of several (unidirectional) couplings.

Additionally, the model structure introduced above should be compared with multi-strain models reviewed in~\citep{Kucharski2016} which do not consider coupled infectious compartments and instead stratify the population by infection history or immunity status. In history-based models for example, individuals that have recovered from infection are returned to one of several susceptible compartments S$_{X}$, where the subscript $X \subseteq \mathcal{N}$ denotes a subset of strains $\mathcal{N} = \left\{1,\ldots,n\right\}$ (of which there are $2^n$) with which the individual has been previously infected.\footnote{For example, individuals with no prior infections are initially introduced into the susceptible compartment S$_{\emptyset}$ whereas an individual who has experienced infection with strains 2 and 5 will return to S$_{2,5}$ upon recovery.} This allows one to incorporate immunological differences that have arisen through past exposure to a particular pathogen strain such as reduced susceptibility to reinfection and/or reduced infectivity. A drawback of this approach is that as the number of strains increases these models quickly become intractable and only models incorporating simplifying assumptions (e.g. complete symmetry between the epidemiological parameters) can be considered for any progress to be made~\citep{Kucharski2016}.

Our model is fundamentally different, in that individuals can transition directly between the infectious compartments without first returning to a susceptible (S) or recovered (R) state. In this case, the number of model parameters scales quadratically (because of $\tens{G}$) as opposed to the exponential scaling of history-based models.\footnote{Conversely, under the assumption of polarized immunity, the number of parameters in status-based models only scales linearly.} Moreover, we do not consider other effects, e.g. cross-immunity, and assume that a) all individuals in compartment S are equally susceptible to infection with any strain and b) all individuals infected with a particular strain $i$ are equally infectious, regardless of their infection history. However, we stress that the $n$-strain model presented in equations~\eqref{eq:dS}-\eqref{eq:dRi} contains parameters which are all strain dependent, thus allowing many of the epidemiological differences arising from past infection to be incorporated. 

\section{System bounds}
\label{sec:system_bounds}

Given the system~\eqref{eq:dS}-\eqref{eq:dRi}, we can show that the region
\begin{equation}
\Omega = \left\{(S,I_i,R_i) \in \mathbb{R}^{2n + 1}_{\geq 0}\; \bigg|  \; 0 \leq N(t) \leq \frac{\lambda}{\mu} \right\}\nonumber
\end{equation}
is a positively invariant and absorbing set that attracts all solutions of~\eqref{eq:dS}-\eqref{eq:dRi} in $\mathbb{R}_{\geq 0}^{2n + 1}$. To see this, we follow the proof outlined in~\citep{bowong2011stability}, whereby we introduce the Lyapunov function $W(t) = S(t) + \sum_{i = 1}^n \left(I_i(t) + R_i(t)\right)$. Taking the time derivative of $W(t)$ and substituting in~\eqref{eq:dS}-\eqref{eq:dRi} we then have
\begin{equation}
\frac{dW}{dt} = \lambda - \mu W - \sum_{i = 1}^n \phi_i I_i \leq \lambda - \mu W.\label{eq:dW}
\end{equation}
For $W > \lambda/\mu$ we have that $\frac{dW}{dt} \leq 0$ which implies that $\Omega$ is a positively invariant set. Moreover, if we solve the inequality~\eqref{eq:dW} we get
\begin{equation}
0 < W(t) \leq \frac{\lambda}{\mu}\left(1 - e^{-\mu t}\right) + W(0)e^{-\mu t},\nonumber
\end{equation}
where the initial condition $W(0) > 0$. Taking the limit as $t\rightarrow \infty$ we have that $0 \leq W(t) \leq \lambda/\mu$ meaning that $\Omega$ is also an attractive set. 

These results imply that the system solutions $(S(t),I_i(t),R_i(t))$ are bounded and nonnegative. 
Moreover, it is straightforward to see that the disease cannot eradicate the population since $N(t) > 0$ for all $t > 0$ and $N(0) \geq 0$. Finally, in the absence of the disease, the population $N(t) \rightarrow S_0 \equiv \lambda/\mu $ as $t\rightarrow\infty$.

\section{Basic reproduction number}
\label{sec:R0}

The basic reproduction number, $R_0$, is defined as the average number of newly infected individuals generated by a single typical infectious individual introduced into a fully susceptible population. In other words, $R_0$ is the initial growth factor (on a generational basis) of infected individuals following the release of an infectious agent. In this sense, $R_0$ represents an important threshold parameter that determines the fate of the infection: for $R_0 > 1$ the disease spreads, and an epidemic occurs; whereas for $R_0 \leq 1$ the infection dies out before a significant fraction of the population is affected.\footnote{We should point out that whilst this is generally true for deterministic, mass-action epidemic models like the one considered here, in stochastic settings for example, there is always a positive probability that the infection dies out quickly, even if $R_0 \geq 1$.}

In the context of multi-strain epidemic models we anticipate a \textit{set} of basic reproduction numbers $R_{0i}$, with each member of the set being associated with the reproductive capacity of each pathogen strain $i$. In this case an epidemic may occur if \textit{any} of the $R_{0i} > 1$. 

To calculate the basic reproduction number associated with each pathogen strain, we follow the procedure outlined in~\citep{Diekmann1990,DandH}. First, we linearize the infected subsystem~\eqref{eq:dImat} about the infection-free equilibrium point, $P_0 = (S_0,\vec{0},\vec{0})$ where $S_0$ is the magnitude of the susceptible population at the infection-free equilibrium, i.e. $S_0 = \lambda/\mu$. This yields the Jacobian, $\bar{\tens{J}}_0$:
\begin{equation}
\bar{\tens{J}}_0 = S_0\tens{B} -\tens{\Delta} + \tens{G}\tens{A}.\nonumber
\end{equation}
We then decompose $\bar{\tens{J}}_0$ into a \textit{transmission} component $\tens{T}$ and a \textit{transition} component $\tens{Q}$, where
\begin{equation}
\tens{T} = S_0\tens{B} \quad \mbox{and} \quad \tens{Q} = \tens{\Delta} - \tens{G}\tens{A},\nonumber
\end{equation}
such that $\bar{\tens{J}}_0 = \tens{T} - \tens{Q}$. This set of matrices is used to calculate the next-generation matrix $\tens{K}$ whose $(j,i)$-th element gives the number of new individuals produced in I$_j$ by those currently in I$_i$. Specifically, we multiply the transmission matrix $\tens{T}$ by the inverse of the transition matrix $\tens{Q}^{-1}$ (whose $(j,i)$-th element measures the mean sojourn time in compartment I$_j$ for an individual currently in I$_i$) to yield
\begin{equation}
\tens{K} = S_0\tens{B}\left(\tens{\Delta} - \tens{G}\tens{A}\right)^{-1} = S_0\tens{B}\tens{\Delta}^{-1}\left(\tens{E} - \tens{G}\tens{A}\tens{\Delta}^{-1}\right)^{-1}.\label{eq:nextgenmat}
\end{equation}
The set of basic reproduction numbers $R_{0i}$ are then given by the spectrum of the next-generation matrix $\mathrm{sp}\left(\tens{K}\right)$ (see~\cite{DandH}).\footnote{Here we use the notation $\mathrm{sp}(\tens{M})$ to denote the set of eigenvalues of the matrix $\tens{M}$.} Following our previous definitions of $\tens{B}$, $\tens{\Delta}$, $\tens{G}$ and $\tens{A}$ (see~\eqref{eq:matG} and~\eqref{eq:mat_def}) we observe that $\tens{K}$ is an $n\times n$ lower triangular matrix. In this case we can simply read off the basic reproduction numbers $R_{0i}$ as the diagonal elements of $\tens{K}$:
\begin{equation}
R_{0i} = \mathrm{sp}(\tens{K})_i = \mathrm{diag}(\tens{K})_i = \mathrm{diag}\left(S_0\tens{B}\tens{\Delta}^{-1}\right)_i.\nonumber
\end{equation}

In component form, the basic reproduction number of the $i$-th strain is given by
\begin{equation}
R_{0i} = \frac{\beta_iS_0}{\delta_i} = \frac{\beta_i S_0}{\mu + \phi_i + \sigma_i + \alpha_i},\label{eq:R0i}
\end{equation}
which we find is purely a function of the epidemiological parameters of strain $i$, and completely independent of the amplification rates $g_{ji}\alpha_i$. This is an important result that shows that mutation events that can continually ignite new strains into existence do not contribute to their reproductive capacity. Although this result is reliant on our assumption of unidirectional flow between the infectious compartments, we believe this to be a reasonable description of the evolutionary process.

Before continuing, we wish to establish the connection between the form of the next-generation matrix $\tens{K}$, equation~\eqref{eq:nextgenmat}, and the infectious compartment network structure discussed in section~\ref{sec:Inetwork}. To do so, we first observe that $\tens{K}$ can be rewritten as
\begin{equation}
\tens{K} = \tens{R}_0\tens{P}\label{eq:KR}
\end{equation}
where $\tens{R}_0 = \mathrm{diag}(R_{0i})$ and $\tens{P}$ is given by
\begin{equation}
\tens{P} = \left(\tens{E} - \tens{G}\tens{A}\tens{\Delta}^{-1}\right)^{-1}.\label{eq:Pdef}
\end{equation}
Immediately we recognize the matrix $\tens{P}$ as the direct analogue of the connectivity matrix~\eqref{eq:con_mat} discussed in section~\ref{sec:Inetwork}, appropriately adjusted to incorporate the rate parameters, $g_{jk} \rightarrow g_{jk}\alpha_k/\delta_k$. Indeed, the elements of $\tens{P}$ can be written in terms of a generating function completely analogous to equation~\eqref{eq:pinfgen}:
\begin{equation}
p_{ji} = \sum_{k = i}^n \frac{g_{jk}\alpha_k}{\delta_k}\,p_{ki}\label{eq:pgen}
\end{equation}
where the initial value $p_{ii} = 1$. Accordingly, the $(j,i)$-th element of the matrix $\tens{P}$ gives the probability of an individual reaching the infectious compartment I$_j$ from compartment I$_i$.

Written in the form~\eqref{eq:KR}, the interpretation of the next-generation matrix becomes even clearer: it is the product of the reproductive capacity of each strain $(\tens{R}_0)$ and the probability of acquiring each strain \textit{once infected} $(\tens{P})$. 
As expected, in the limit $\tens{G}\rightarrow\tens{0}$, the matrix $\tens{P}$ reduces to the identity matrix $\tens{E}$ and we recover the canonical result derived for uncoupled systems (see section~\ref{sec:un_comp}).

\section{Epidemic outbreak}
\label{sec:epidemic_outbreak}

We now return to the statement made at the opening of the previous section: if each of the $R_{0i} \leq 1$, the diseased population quickly fades out; however, if $R_{0i} > 1$ for any $i$ an epidemic occurs. To see this we simply rewrite equation~\eqref{eq:dIi} in terms of the basic reproduction numbers $R_{0i}$:
\begin{equation}
\frac{dI_i}{dt} = \delta_i\left(R_{0i}\frac{S}{S_0} - 1\right)I_i + \sum_{j = 1}^{i-1} g_{ij}\alpha_j I_j.
\end{equation}
Taking as our initial condition $S(0) = S_0$ (which is the most optimistic case from the disease's perspective), we can clearly see that if $R_{0i} \leq 1$ for all $i$, $I_i(t)$ decays exponentially and the disease rapidly fades out (see next section for a more rigorous analysis). Conversely, if $R_{0i} > 1$, $I_i(t)$ grows exponentially at early times and the epidemic takes off. Therefore, the basic reproduction numbers $R_{0i}$ represent a set of threshold parameters that determine the fate of the infectious agent.

In fact, if we introduce the notation
\begin{equation}
R_{0f} \equiv \max_i \, R_{0i}\label{eq:R0f}
\end{equation}
to single out the strain the greatest reproduction number, or the fittest strain $f$, we can restate this result more succinctly as: an epidemic occurs when $R_{0f} > 1$ ,whereas the disease dies out when $R_{0f} \leq 1$. 

We point out that here, we have implicitly assumed that fittest strain $f$ is \textit{observable}, i.e. that for some time $t \geq 0$, $I_f(t)$ is positive. In our model, this can be achieved if either $I_f(0) > 0$ \textit{or} $I_j(0) > 0$ for some $j\triangleleft f$. That is, the initial seed that sparks the exponential growth of $I_f(t)$ can come from strain $f$, or any one of strain $f$'s ancestors. By the same token, we are compelled to augment our definition of the fittest strain $f$ to ensure that it is in fact observable, which will be assumed from here out. Nevertheless, in this case, the space of initial conditions that lead to an epidemic outbreak is in fact larger than that in the corresponding uncoupled multi-strain model, which strictly requires the initial population of strain $f$ itself to be positive. 


\section{Asymptotic system dynamics}
\label{sec:dynamics}

Having discussed the initial epidemic trajectory in terms of the basic reproduction numbers, $R_{0i}$, we now turn our attention to the late-time dynamics of the system. First, we show (by induction) that if $R_{0f} \leq 1$ ($\Leftrightarrow R_{0i} \leq 1$ for all $i$) the disease is eradicated from the population. We will then consider the alternative case $R_{0f} > 1$ and show that only the fittest strain $f$, along with each of its descendants $j\triangleright f$, remain in circulation indefinitely. The general approach used to arrive at each of these conclusions is based upon the analysis in~\citep{Bremermann}.

For each case, we introduce the bound
\begin{equation}
\limsup\limits_{t\rightarrow\infty} N(t) \leq \frac{\lambda}{\mu} = S_0\nonumber
\end{equation}
which follows immediately from the invariance of $\Omega$.

\subsection{Case 1: $R_{0f} \leq 1$}

For the case $R_{0f} \leq 1$, we choose a sequence $t_m\rightarrow\infty$ for $m\rightarrow\infty$ such that $I_i(t_m)\rightarrow \limsup\limits_{t\rightarrow\infty} I_i(t)$ and $\dot{I}_i(t_m) \geq 0$. From equations~\eqref{eq:Nsum} and~\eqref{eq:dIi} we then have
\begin{equation}
0 \leq \limsup_{t\rightarrow\infty} I_i(t) \left(\frac{\beta_i}{\delta_i}\left(S_0 - \limsup\limits_{t\rightarrow\infty} I_i(t)\right) - 1\right) + \sum_{j = 1}^{i-1} \frac{g_{ij}\alpha_j}{\delta_i} \limsup\limits_{t\rightarrow\infty} I_j(t).\nonumber
\end{equation}
To show that this expression does in fact imply that $\lim_{t\rightarrow\infty} I_i(t) = 0$ for all $i$ when $R_{0f} \leq 1$ we start with strain $i=1$, for which the amplification term vanishes and we are left with\footnote{For the uncoupled case considered in~\citep{Bremermann} the amplification term is absent for all $i$ and the conclusion $\lim_{t\rightarrow\infty} I_i(t) = 0$ for $R_{0f} \leq 1$ follows immediately from inspection.} 
\begin{equation}
0 \leq \limsup_{t\rightarrow\infty} I_1(t) \left(\frac{\beta_1}{\delta_1}\left(S_0 - \limsup\limits_{t\rightarrow\infty} I_1(t)\right) - 1\right).\nonumber
\end{equation}
Hence, if 
\begin{equation}
R_{01} = \frac{\beta_1 S_0}{\delta_1} \leq 1 \quad \Rightarrow \quad \limsup\limits_{t\rightarrow\infty} I_1(t) = \lim\limits_{t\rightarrow\infty} I_1(t) = 0\label{eq:I1vanish}
\end{equation}
where we have used the fact that $\liminf\limits_{t\rightarrow\infty} I_i(t) \geq 0$ which also follows from the invariance of $\Omega$. 

Continuing, for strain 2 we then have
\begin{equation}
0 \leq \limsup_{t\rightarrow\infty} I_2(t) \left(\frac{\beta_2}{\delta_2}\left(S_0 - \limsup\limits_{t\rightarrow\infty} I_2(t)\right) - 1\right)\nonumber
\end{equation}
where again the contribution from the amplification term vanishes as a consequence of~\eqref{eq:I1vanish}. Similarly, we find that if $R_{02} \leq 1$, $\lim_{t\rightarrow\infty} I_2(t) = 0$. 

By induction we can show that all strains $i$ are driven to extinction (i.e. $\lim_{t\rightarrow\infty} I_i(t) = 0$ for all $i$) provided $R_{0f} \leq 1$.

In a similar manner it is possible to show that the recovered populations $R_i(t)$ are also driven to zero when $R_{0f}\leq 1$ and that 
(see equation~\eqref{eq:dS}) $\lim_{t\rightarrow\infty} S(t) = \lambda/\mu = S_0$.

In fact, we can also show that the infection-free equilibrium point $P_0 = (S_0, \vec{0}, \vec{0})$ is locally asymptotically stable if $R_{0f} < 1$. To prove this we consider the Jacobian of the system~\eqref{eq:dS}-\eqref{eq:dRi} which is given by
\begin{equation}
\tens{J} = 
\begin{bmatrix}
-\vec{\beta}^\intercal\vec{I} - \mu & -S\vec{\beta} + \rho\vec{\eta} & \vec{\gamma} \\
\tens{B}\vec{I}  & S\tens{B} - \tens{\Delta} + \tens{G}\tens{A} & \tens{0} \\
\vec{0} & (1 - \rho)\tens{H} & -\tens{M} - \tens{\Gamma} \\
\end{bmatrix}.\nonumber
\end{equation}
At the infection-free equilibrium point, $P_0$, this reduces to\footnote{Note that the Jacobian of the infected subsystem $\bar{\tens{J}}_0$ introduced in section~\ref{sec:R0} is a sub-matrix of $\tens{J}_0$.}
\begin{equation}
\tens{J}_0 = 
\begin{bmatrix}
- \mu & -S_0\vec{\beta} + \rho\vec{\eta} & \vec{\gamma} \\
\vec{0}  & S_0\tens{B} - \tens{\Delta} + \tens{G}\tens{A} & \tens{0} \\
\vec{0} & (1 - \rho)\tens{H} & -\tens{M} - \tens{\Gamma} \\
\end{bmatrix}.\label{eq:J0}
\end{equation}
The structure of $\tens{J}_0$ allows us to immediately read off the $2n + 1$ eigenvalues, $\theta$, as
\begin{equation}
\theta_0 = -\mu < 0,\quad \theta_{i} = -\mu - \gamma_i < 0,\nonumber
\end{equation}
and
\begin{equation}
\theta_{n + i} = \mathrm{sp}\left(S_0\tens{B} - \tens{\Delta} + \tens{G}\tens{A}\right)_i = \mathrm{diag}\left(S_0\tens{B} - \tens{\Delta} + \tens{G}\tens{A}\right)_i = \delta_i\left(R_{0i} - 1\right).\nonumber
\end{equation}
where $i \in [1,n]$. Hence, if $R_{0f} < 1$ ($\Leftrightarrow R_{0i} < 1$ for all $i$) the infection-free equilibrium point $P_0$ is locally asymptotically stable. Conversely, if $R_{0i} > 1$ for any $i$, the Jacobian~\eqref{eq:J0} has at least one positive eigenvalue and $P_0$ is unstable.


\subsection{Case 2: $R_{0f} > 1$}

We now consider the asymptotic behaviour of the system in the alternative case, when $R_{0f} > 1$. As a first step, we solve~\eqref{eq:dIi} using the variation of constants method to get
\begin{align}
I_i(t) &= I_i(0)\exp \left[\beta_i \int_{0}^t S(t')\,dt' - \delta_i t\right] \nonumber\\
&\qquad  + \sum_{j = 1}^{i-1} g_{ij}\alpha_j \int_{0}^t I_j(t - t') \exp \left[\beta_i\int_{t - t'}^t S(t'')\, dt'' - \delta_i t'\right]\,dt'.\label{eq:I_gensol}
\end{align}
From this we can see that the exponential growth or decay of each infectious population $I_i(t)$ depends on an integral of the form (c.f. the discussion at the end of section~\ref{sec:epidemic_outbreak})
\begin{equation}
\int \left(\beta_i S(t') - \delta_i\right) dt'.\nonumber
\end{equation}
Substituting in the definition of $R_{0i}$, and keeping in mind that solutions of the system~\eqref{eq:dS}-\eqref{eq:dRi} are bounded (see section~\ref{sec:system_bounds}), it follows that the average asymptotic value of the susceptible population $S(t)$ must be bounded from above:
\begin{equation}
\lim_{t\rightarrow\infty} \frac{1}{t}\int_0^t S(t')\,dt' \leq \frac{S_0}{R_{0i}} , \qquad \forall \: i.\nonumber
\end{equation}

Combining this result with the restriction $S(t) \leq S_0$, we therefore conclude that
\begin{equation}
\lim_{t\rightarrow\infty} \frac{1}{t}\int_0^t S(t')\,dt'  \leq \min_{i}\left[S_0,\frac{S_0}{R_{0i}}\right] = \min\left[S_0,\frac{S_0}{R_{0f}}\right].\label{eq:Savg}
\end{equation}
The simple case $R_{0f} \leq 1$, which we have already considered, is consistent with $\lim_{t\rightarrow\infty} S(t) = S_0$. However, for the alternative case $R_{0f} > 1$, the average value of the susceptible population resides below the infection-free upper bound. Consequently, we shall see that several circulating strains will be driven to extinction because they cannot be sustained by the diminished susceptible pool.

To show this, we again begin with strain $i=1$, for which~\eqref{eq:I_gensol} reduces to
\begin{equation}
I_1(t) = I_1(0)\exp \left(\beta_1 \int_{0}^t S(t')\,dt' - \delta_1 t\right). \nonumber
\end{equation}
Assuming $f \neq 1$ so that $R_{01} < R_{0f}$, the condition~\eqref{eq:Savg} ensures that $I_1(t)$ decays exponentially in the limit $t\rightarrow\infty$, i.e. $\lim\limits_{t\rightarrow\infty} I_1(t) = 0$. 

Similarly, when $i = 2$, the general solution~\eqref{eq:I_gensol} becomes
\begin{align}
I_2(t) &= I_2(0)\exp \left(\beta_2 \int_{0}^t S(t')\,dt' - \delta_2 t\right) \nonumber\\
&\qquad  + g_{21}\alpha_1 \int_{0}^t I_1(t - t') \exp \left(\beta_2\int_{t - t'}^t S(t'')\, dt'' - \delta_2 t'\right)\,dt'.\nonumber\\
&= \exp \left(\beta_2 \int_{0}^t S(t')\,dt' - \delta_2 t\right)\left[I_2(0) + g_{21}\alpha_1 \int_0^t I_1(t') \exp \left(-\beta_2 \int_{0}^{t'} S(t'')\,dt'' + \delta_2 t'\right)dt'\right]\nonumber
\end{align}
where in the second line we have factored out the exponential $\exp \left(\beta_2 \int_{0}^t S(t')\,dt' - \delta_2 t\right)$ from each term. Again, assuming this time $f > 2$, we can use~\eqref{eq:Savg} and the fact that both terms in the square brackets are bounded to see that $\lim\limits_{t\rightarrow\infty} I_2(t) = 0$. By induction, it is straightforward to show that the remaining non-descendants of strain $f$ also satisfy $\lim_{t\rightarrow\infty} I_i(t) = 0$.

The next task is to show that strain $f$ persists provided $R_{0f} > 1$. To do this, we first show that the inequality in~\eqref{eq:Savg} is in fact a strict equality using a proof by contradiction. 

Firstly, let us take the case $R_{0f} > 1$ and assume that the inequality in~\eqref{eq:Savg} is strict, such that
\begin{equation}
\lim_{t\rightarrow\infty} \frac{1}{t}\int_0^t S(t')\,dt'  < \frac{S_0}{R_{0f}}.\nonumber
\end{equation}
Substituting this condition into~\eqref{eq:I_gensol} we then have that $\lim_{t\rightarrow\infty} I_i(t) = 0$ for all $i$. However, from equation~\eqref{eq:dS}, it must then follow that $\lim_{t\rightarrow\infty} S(t) = S_0$: a contradiction. Therefore, in the case $R_{0f} > 1$, we must have
\begin{equation}
\lim_{t\rightarrow\infty} \frac{1}{t}\int_0^t S(t')\,dt'  = \frac{S_0}{R_{0f}},\nonumber
\end{equation}
or, in general
\begin{equation}
\lim_{t\rightarrow\infty} \frac{1}{t}\int_0^t S(t')\,dt'  = \min\left[S_0,\frac{S_0}{R_{0f}}\right].
\end{equation}

As a corollary, we can see that in the case $R_{0f} > 1$, $\liminf\limits_{t\rightarrow\infty} I_f(t) = \epsilon > 0$, meaning that strain $f$ persists asymptotically. Equation~\eqref{eq:I_gensol} can then be used to show that all $j\triangleright f$ survive asymptotically as well.



The critical point in the preceding analysis about the limited availability of susceptibles is a hallmark of the competitive exclusion principle: when several species are competing over the same shared resource --- in this case, susceptibles --- only the one with the greatest reproduction number can survive indefinitely; the remaining species are driven to extinction. However, in the coupled case considered here, in which circulating strains are genetically related via the amplification mechanism previously described, we must extend this principle to include all those strains that are descendants of the fittest strain $f$ (in the sense defined in section~\ref{sec:Inetwork}) among those that survive indefinitely; here, each mutant descendant is perpetually fueled by the fittest strain.


\section{Equilibrium solutions}
\label{sec:equilibrium}

\begin{figure}[htbp]
\centering
\includegraphics[width=0.7\textwidth]{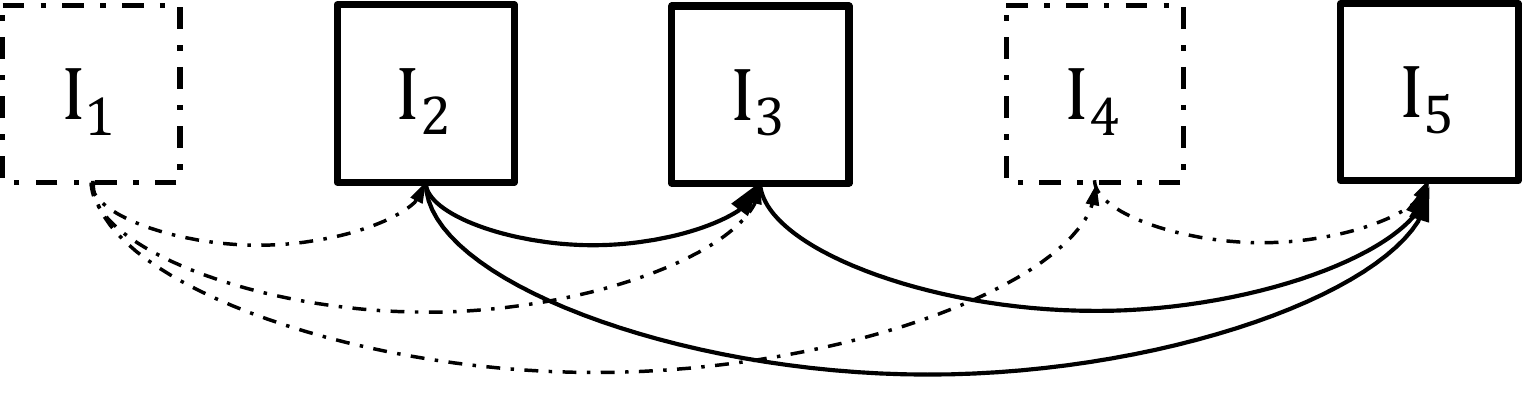}
\caption{Example of an incomplete I$_i$ flow network at an endemic equilibrium point. Here we have chosen $n = 5$ strains and considered the endemic equilibrium point $P^{*2}$ at which the second strain becomes to source strain. Solid boxes and arrows indicate those compartments and flows that persist at equilibrium whilst dot-dashed lines correspond to those that have been driven to extinction. Notice that only strain 2 and its descendants, strains 3 and 5, survive.} 
\label{fig:Inetwork_equil}
\end{figure}

To reinforce the asymptotic analysis presented in the previous section, here we determine the equilibrium states of the system~\eqref{eq:dS}-\eqref{eq:dRi}. In particular, we find that in addition to the infection-free equilibrium point $P_0 = (S_0,\vec{0},\vec{0})$, there are a set of $n$ endemic equilibria, $P^{*i} = (S^{*i}, \vec{I}^{*i} \neq \vec{0},\vec{R}^{*i}\neq 0)$ where $i\in[1,n]$. Explicitly, we will show that at the $i$-th endemic equilibrium point $P^{*i}$, all strains are driven to extinction 
except for strain $i$ and its $d_i$ descendants.\footnote{Recall from section~\ref{sec:Inetwork} that the number of descendants of strain $i$, denoted $d_i$ is equal to the number of non-zero elements in the $i$-th column of the connectivity matrix $\tens{P}_\infty$~\eqref{eq:con_mat}, or, when updated to incorporate the rate parameters $\alpha_k$ and $\delta_k$, the matrix $\tens{P}$~\eqref{eq:Pdef}.} Nominally, at the $i$-th endemic equilibrium, strain $i$ becomes the \textit{source} strain (i.e. all of the ``upstream" strains have died out) and the number of strains in circulation is equal to $d_i + 1$ (see figure~\ref{fig:Inetwork_equil}). However, despite there being the possibility of $n$ endemic equilibrium solutions, we find that only a limited subset of these lie within $\mathbb{R}_{\geq 0}^{2n+1}$, i.e. only a limited subset are physical. We will show that the absolute and relative magnitudes of the basic reproduction numbers of each strain regulate the existence of physical solutions.

Lastly, we demonstrate that the fittest strain is not necessarily the most prevalent, and that a descendant strain can outnumber the source strain.

To calculate the equilibrium states of the system we begin by setting the left-hand side of~\eqref{eq:dSmat}-\eqref{eq:dRmat} to zero:
\begin{align}
0 &= \lambda - \mu S - S\vec{\beta}^\intercal \vec{I} + \rho\vec{\eta}^\intercal \vec{I} + \vec{\gamma}^\intercal\vec{R},\label{eq:dSmat_equil}\\
\vec{0} &= \left(S\tens{B} - \tens{\Delta} + \tens{G}\tens{A}\right)\vec{I},\label{eq:dImat_equil}\\
\vec{0} &= (1 - \rho)\tens{H}\vec{I} - (\mu\tens{E} + \tens{\Gamma})\vec{R}.\label{eq:dRmat_equil}
\end{align}
We shall focus initially on the second equation~\eqref{eq:dImat_equil}, which, as noted previously, is invariant between the SIS, SIR and SIRS models. 


The infection-free equilibrium solution $\vec{I}_0 = \vec{0}$ has been discussed in the previous section. To find the endemic solutions, $P^{*i}$, for which $(\vec{I}^{*i} \neq \vec{0})$ we first pre-multiply equation~\eqref{eq:dImat_equil} through by $(S^{*i}\tens{B})^{-1}\tens{K}$ to get
\begin{equation}
\vec{0} = \left(\tens{B}^{-1}\tens{K}\tens{B} - \frac{S_0}{S^{*i}}\tens{E}\right)\vec{I}^{*i},\label{eq:Ieigen}
\end{equation}
where we have substituted in the definition of the next-generation matrix~\eqref{eq:nextgenmat}. Here we recognize equation~\eqref{eq:Ieigen} as the eigenvalue equation for the matrix $\tens{B}^{-1}\tens{K}\tens{B}$. Accordingly, we may identify the quantities $S_0/S^{*i}$ as the eigenvalues of $\tens{B}^{-1}\tens{K}\tens{B}$ and the solutions $\vec{I}^{*i}$ as the corresponding eigenvectors. Therefore, the number of endemic solutions of the system~\eqref{eq:dS}-\eqref{eq:dRi} equals the number of distinct eigenvalues of $\tens{B}^{-1}\tens{K}\tens{B}$.

Given that $\tens{B}^{-1}\tens{K}\tens{B}$ is a similarity transform of the next-generation matrix $\tens{K}$, 
 their eigenvalues are identical, allowing us to write
\begin{equation}
\frac{S_0}{S^{*i}} = \mathrm{sp}\left(\tens{B}^{-1}\tens{K}\tens{B}\right)_i = \mathrm{sp}\left(\tens{K}\right)_i = R_{0i}.\nonumber
\end{equation}
Hence, the size of the susceptible population at the $i$-th endemic equilibrium point $P^{*i}$ assumes its canonical value derived for the uncoupled multi-strain case (see next section),
\begin{equation}
S^{*i} = \frac{S_0}{R_{0i}}.\label{eq:Si}
\end{equation}

The size of the infected population at each endemic equilibrium point $P^{*i}$ can be determined by finding the set of eigenvectors of the lower triangular matrix $\tens{B}^{-1}\tens{K}\tens{B}$. In this case, solving equation~\eqref{eq:Ieigen} alone will only allow us to calculate the solutions $\vec{I}^{*i}$ up to a multiplicative constant $\vec{v}_i \propto \vec{I}^{*i}$; to determine the normalization factors we will need to use~\eqref{eq:dSmat} and~\eqref{eq:dRmat}.
 
The elements of each eigenvector solution $\vec{I}^{*i}$ can be calculated most easily by returning to the component form equation for $dI_j/dt$ (note that we now use $j$ as the free variable to label the various strains since the superscript $i$ labels the endemic solution):
\begin{equation}
0 = \delta_j\left(\frac{R_{0j}}{R_{0i}} - 1\right)I_j^{*i} + \sum_{k = 1}^{j-1} g_{jk}\alpha_k I_k^{*i},\label{eq:Ireczero}
\end{equation}
where we have substituted in~\eqref{eq:Si} and the definition of the basic reproduction number~\eqref{eq:R0i}. 

By studying~\eqref{eq:Ireczero} we can see that the first non-zero value in the sequence of $I_j^{*i}$, is $I_i^{*i}$ itself.\footnote{To see this, consider the first equation in the sequence ($j=1$) for which the summation term vanishes and we have
\begin{equation}
0 = \delta_1\left(\frac{R_{01}}{R_{0i}} - 1\right)I_1^{*i}.
\end{equation}
Assuming $i \neq 1$, we must have that $I_1^{*i} = 0$, since $R_{0i} \neq R_{0j}$ for all $i \neq j$. In turn, from the equation for $j = 2$, where again the summation term $\sum_{k = 1}^{1}g_{2k}\alpha_{k}I_k^{*i} = g_{21}\alpha_1I_1^{*i} = 0$, we similarly find $I_2^{*i} = 0$ (assuming $i \neq 2$). Progressing sequentially through each of the $I_j^{*i}$ we find that they are all equal to zero until eventually we reach $j = i$ . Here, the coefficient of $I_i^{*i}$ is zero, permitting $I_i^{*i} \neq 0$ solutions. (If $I_i^{*i}$ did equal zero then so would the remaining $I_j^{*i}$ for $j > i$.)
} This means that all strains upstream from strain $i$ (that is, all $j < i$) die out at the equilibrium point $P^{*i}$. 

In this case, the summation index $k$ (partially) collapses and we have
\begin{equation}
I_j^{*i} = \frac{R_{0i}}{R_{0i} - R_{0j}}\sum_{k = i}^{j - 1} \frac{g_{jk}\alpha_k}{\delta_j} I_k^{*i},\label{eq:Irec}
\end{equation}
where the initial value $I_i^{*i}$ will be determined below.

First, we observe that the recursive equation~\eqref{eq:Irec} is of exactly the same form as the generating function for the connectivity matrix~\eqref{eq:pinfgen}, albeit with the mutation rates $g_{jk}$ replaced by
\begin{equation}
g_{jk} \rightarrow \frac{R_{0i}}{R_{0i} - R_{0j}} \frac{g_{jk}\alpha_k}{\delta_j}.
\end{equation}
Therefore, following our discussion in section~\ref{sec:Inetwork}, we can see that the only non-vanishing elements in the series $I_j^{*i}$ are those which correspond to the infectious compartments I$_j$ that are connected to compartment I$_i$. 
Therefore, at the $i$-th endemic equilibrium point $P^{*i}$ only strain $i$ and its descendants remain in circulation.
 
We can also use the recursive expression~\eqref{eq:Irec} to determine the sign of each of the $I_j^{*i}$ relative to $I_i^{*i}$ to help ascertain whether the equilibrium point $P^{*i}$ lies in the non-negative orthant $\mathbb{R}_{\geq 0}^{2n+1}$; or, in other words, whether the equilibrium point $P^{*i}$ is physical. We will then determine the sign of $I_i^{*i}$ itself using~\eqref{eq:Ndot}. 

Using~\eqref{eq:Irec} and progressing through each $I_j^{*i}$ for $j > i$, we observe that a necessary and sufficient condition for $\mathrm{sgn} (I_j^{*i}) = \mathrm{sgn} (I_i^{*i})$ is to have $R_{0i} > R_{0j}$ for all $j \triangleright i$. That is, $R_{0i}$ must maximize the set of basic reproduction numbers among each of its descendants. Therefore, the nonnegativity of the endemic solutions depends on the relative magnitudes of the basic reproduction numbers of strain $i$ and its descendants $j$. If the condition $R_{0i} > R_{0j}$ is violated for any $j \triangleright i$ then there must be a sign change between $I_i^{*i}$ and $I_j^{*i}$ which automatically signifies that $P^{*i}$ lies outside $\mathbb{R}_{\geq 0}^{2n+1}$ and is therefore non-physical.

Finally, to determine the sign of $I_i^{*i}$ we use equation~\eqref{eq:Ndot}, which at the $i$-th endemic equilibrium point becomes
\begin{equation}
0 = \lambda - \mu N^{*i} - \sum_{j = 1}^n \phi_j I_j^{*i}.
\end{equation}
Substituting in the identities $N^{*i} = S^{*i} + \sum_{j = 1}^{i}\left(I_j^{*i} + R_j^{*i}\right)$, $S^{*i} = S_0/R_{0i}$ along with the solution of equation~\eqref{eq:dRmat_equil}, $R_j^{*i} = (1-\rho)\eta_j I_j^{*i}/(\mu + \gamma_j)$, and rearranging we find
\begin{equation}
\sum_{j = 1}\left[\mu + \phi_j + \left(1 - \rho\right)\frac{\mu\eta_j}{\mu + \gamma_j}\right]I_j^{*i} = \mu S_0\left(1 - \frac{1}{R_{0i}}\right).\label{eq:Inorm}
\end{equation}
By using equation~\eqref{eq:Irec} we observe that each term in the summation on the left-hand side of~\eqref{eq:Inorm} is proportional to $I_i^{*i}$, and, furthermore, that the coefficients are all positive provided $R_{0i} > R_{0j}$ for all $j\triangleright i$. If these conditions are satisfied we then have
\begin{equation}
\mathrm{sgn}\left(I_i^{*i}\right) = \mathrm{sgn}\left(1 - \frac{1}{R_{0i}}\right).\label{eq:signresult}
\end{equation}
Therefore, if $R_{0i} > 1$, $I_i^{*i}$ is positive; if $R_{0i} < 1$, $I_i^{*i}$ is negative. In the special case $R_{0i} = 1$, the endemic equilibrium point $P^{*i}$ coincides with the infection-free equilibrium $P_0$. Combining this with our earlier result we find that the necessary and sufficient conditions for $P^{*i}$ to lie in $\mathbb{R}_{\geq 0}^{2n +1}$ are to have $R_{0i} \geq 1$ and $R_{0i} > R_{0j}$ for all $j \triangleright i$. 

In support of the analysis presented in the previous section we have now found that when $R_{0f} \leq 1$ the infection-free equilibrium point $P_0$ is the unique fixed point in $\Omega$. Conversely, when $R_{0f} > 1$, we have found that only those endemic equilibria that simultaneously satisfy the constraints $R_{0i} > 1$ and $R_{0i} > R_{0j}$ for all $j\triangleright i$ are nonnegative. Naturally, the latter condition is automatically satisfied for the endemic equilibrium point $P^{*f}$ associated with the fittest strain $f$ (since $R_{0f} \equiv \max_i \, R_{0i}$).\footnote{The appearance of additional, spurious endemic equilibria in $\Omega$ can be traced back to the discussion about initial conditions in section~\ref{sec:R0}; these points would only be approached asymptotically if strain $f$ was never observed, that is, if $I_f(0) = I_j(0) = 0$ for all $j\triangleleft f$.}




Beyond the non-negativity of the endemic equilibrium solutions we should also investigate the analytic form and relative magnitudes of the various infectious populations at $P^{*i}$. For this purpose we again turn to the recursive relation~\eqref{eq:Irec}. 

Restricting our attention to only those endemic solutions that are physical (i.e. those states for which $R_{0i} \geq 1$ and $R_{0i} > R_{0j}$ for all $j \triangleright i$), inspection of~\eqref{eq:Irec} reveals that a sufficient condition for $I_{j}^{*i} > I_{i}^{*i}$ is
\begin{equation}
\frac{R_{0i}}{R_{0i} - R_{0j}}\frac{g_{ji}\alpha_i}{\delta_{j}} > 1,\label{eq:con2}
\end{equation} 
which can easily be satisfied if $R_{0j}$ is sufficiently close to $R_{0i}$. This, of course, is only one particular example; the condition $I_{j}^{*i} > I_{i}^{*i}$ for some $j \triangleright i$ can be more easily satisfied if we include the extra contribution from the remaining terms in the summation~\eqref{eq:Irec}.

Remarkably though, even if we switch off the direct transmission contribution for all the descendant strains, by taking the limit $\beta_j\rightarrow 0$ for all $j \triangleright i$, the condition $I_{j}^{*i} > I_{i}^{*i}$ can still be achieved if
\begin{equation}
\frac{g_{ji}\alpha_i}{\delta_j} > 1.\nonumber
\end{equation}
Intuitively this condition is straightforward: if the transition rate from I$_i$ to I$_j$ ($g_{ji}\alpha_i$) is greater than the removal rate from compartment I$_j$ ($\delta_j$), the descendant strain $j$ will grow to become more prevalent than the source strain $i$. Therefore, even in the absence of direct transmission, a descendant strain can become more prevalent than the source strain. Keep in mind, this only represents the contribution from a single term in the summation~\eqref{eq:Irec}, so therefore represents a lower bound on the relative magnitude of $I_j^{*i}$; restoring the extra contribution from direct transmission only broadens the parameter range under which the circumstance $I_j^{*i} > I_i^{*i}$ is realized.

Finally, using the normalized factorization for the endemic equilibrium solution $\vec{I}^{*i} = I_i^{*i}\vec{v}_i$ allows us to write the endemic solutions as
\begin{equation}
\vec{I}^{*i} = \frac{\mu \left(R_{0i} - 1\right)\vec{v}_i}{\left[\vec{\beta}^\intercal - \frac{\beta_i}{\delta_i}\vec{\eta}^\intercal\left(\mu\tens{E} + \tens{\Gamma}\right)^{-1}\left(\mu\rho\tens{E} + \tens{\Gamma}\right)\right]\vec{v}_{i}},\label{eq:Ifinal}
\end{equation}
where the normalization factor (i.e. $I_i^{*i}$) in this case has been calculated by rearranging~\eqref{eq:dSmat_equil} rather than using~\eqref{eq:Inorm}.

Below we summarize the endemic solutions for the SIS ($\rho = 1$), SIR ($\rho = 0, \vec{\gamma} = \vec{0}$) and SIRS ($\rho = 0$) coupled epidemic models:

SIS:
\begin{equation}
\vec{I}^{*i}=\frac{\mu\left(R_{0i} - 1\right)\vec{v}_i}{\left(\vec{\beta}^\intercal - \frac{\beta_i}{\delta_i}\vec{\eta}^\intercal\right)\vec{v}_i}.
\end{equation}

SIR: 
\begin{equation}
\vec{I}^{*i}=\frac{\mu\left(R_{0i} - 1\right)\vec{v}_i}{\vec{\beta}^\intercal\vec{v}_i}.
\end{equation}

SIRS:
\begin{equation}
\vec{I}^{*i}=\frac{\mu\left(R_{0i} - 1\right)\vec{v}_i}{\left(\vec{\beta}^\intercal - \frac{\beta_i}{\delta_i}\vec{\eta}^\intercal\left(\mu\tens{E} + \tens{\Gamma}\right)^{-1}\tens{\Gamma}\right)\vec{v}_i}.
\end{equation}

\section{Comparison with uncoupled models}
\label{sec:un_comp}

The expressions derived in the previous two sections for the coupled SIS, SIR and SIRS models are all generalizations of the familiar uncoupled results. In this section we demonstrate how the uncoupled expressions can be derived from the more general expressions given above to retrospectively reveal how the coupling term modifies the behaviour of uncoupled multi-strain models.

To begin, we remind the reader that the uncoupled case can be derived from the general case by taking the limit $\tens{G}\rightarrow \tens{0}$ in the system~\eqref{eq:dSmat}-\eqref{eq:dRmat}. The uncoupled system of equations is then given by
\begin{align}
\frac{dS}{dt} &= \lambda - \mu S - S\vec{\beta}^\intercal \vec{I} + \rho\vec{\eta}^\intercal \vec{I} + \vec{\gamma}^\intercal\vec{R},\label{eq:dSmat_u}\\
\frac{d\vec{I}}{dt} &= \left(S\tens{B} - \tens{\Delta} \right)\vec{I},\label{eq:dImat_u}\\
\frac{d\vec{R}}{dt} &= (1 - \rho)\tens{H}\vec{I} - (\mu\tens{E} + \tens{\Gamma})\vec{R},\label{eq:dRmat_u}
\end{align} 
where most of the variables retain their earlier definitions however, $\eta_i$ (and therefore $\tens{H}$) now simplifies to
\begin{align}
\eta_i &= \sigma_i + \alpha_i,\nonumber\\
\tens{H} &= \mathrm{diag}\left(\vec{\eta}\right) = \mathrm{diag}\left(\vec{\sigma} + \vec{\alpha}\right).
\end{align}

In a similar manner, the expression for the next-generation matrix in the uncoupled case can easily be found by taking the limit $\tens{G} \rightarrow \tens{0}$ in equation~\eqref{eq:nextgenmat} to get
\begin{equation}
\tens{K}_u = S_0\tens{B}\tens{\Delta}^{-1},\nonumber
\end{equation}
where we have introduced a subscript $u$ to denote the uncoupled version of $\tens{K}$. This matrix is strictly diagonal and yields the exact same expressions for the basic reproduction numbers of each strain as derived previously for the general coupled case:
\begin{equation}
R_{0i,u} = \mathrm{sp}\left(\tens{K}_u\right)_i = \frac{\beta_i S_0}{\delta_i} = R_{0i}.\nonumber
\end{equation}
Hence, as mentioned in section~\ref{sec:R0}, the coupling term does not modify the expression for the basic reproduction number $R_{0i}$ of each strain, however it does transform the strictly diagonal next-generation matrix $\tens{K}_u$ into a lower triangular matrix $\tens{K}$. Indeed, we can rewrite the next-generation matrix $\tens{K}_u$ in terms of the diagonal matrix of basic reproduction numbers $\tens{R}_0$: $\tens{K}_u = \tens{R}_0$, where $\tens{R}_0$ is the exact same matrix as the one introduced in section~\ref{sec:R0}.

Continuing on to the equilibrium solutions we very quickly notice that the infection-free equilibrium is unchanged whether or not we include the coupling term (see previous section). However, by removing the coupling between the strains, the endemic equilibrium solutions simplify considerably. To see this, replace the coupled next-generation matrix $\tens{K}$ in~\eqref{eq:Ieigen} with its uncoupled equivalent $\tens{K}_u$, in which case the matrix $\tens{B}^{-1}\tens{K}\tens{B}$ becomes
\begin{equation}
\tens{B}^{-1}\tens{K}\tens{B} \: \rightarrow \: \tens{B}^{-1}\tens{K}_u\tens{B} = \tens{B}^{-1}\tens{B} \tens{K}_u= \tens{K}_u\nonumber
\end{equation}
where we have used the fact that both $\tens{K}_u$ and $\tens{B}$ are diagonal so that $\tens{K}_u\tens{B} = \tens{B}\tens{K}_u$. Therefore, the uncoupled version of the eigenvalue equation~\eqref{eq:Ieigen} is given by
\begin{equation}
\vec{0} = \left(\tens{K}_u - \frac{S_0}{S^{*}}\tens{E}\right)\vec{I}_u^*.\label{eq:Ieigen_u}
\end{equation}
As before, we have
\begin{equation}
\frac{S_0}{S^{*i}} = \mathrm{sp}(\tens{K}_u)_i = R_{0i} \quad \Rightarrow \quad S^{*i} = \frac{S_0}{R_{0i}}.\nonumber
\end{equation}
However, since $\tens{K}_u$ is a diagonal matrix, the uncoupled endemic solutions $\vec{I}^{*i}_u$ are just proportional to the $n$-dimensional canonical basis vectors $\vec{e}_i$ which contain all zeros except for the $i$-th element which is set to unity. Interpreting this physically, it means that at the $i$-th endemic equilibrium point only strain $i$ survives with all other strains driven to extinction.

Going further and substituting $\vec{v}_i\rightarrow \vec{v}_i^u = \vec{e}_i$ into the final solution~\eqref{eq:Ifinal} and rearranging gives
\begin{equation}
\vec{I}^{*i}_u = \frac{\mu}{\beta_i}\left(R_{0i} - 1\right)\left(1 - \frac{\eta_i}{\delta_i}\frac{\mu\rho + \gamma_i}{\mu + \gamma_i}\right)^{-1}\vec{e}_i.\nonumber\\
\end{equation}
The particular solutions for the uncoupled SIR, SIS and SIRS models are summarized below 

SIS:
\begin{equation}
\vec{I}^{*i}_u = \frac{\mu S_0}{\delta_i - \eta_i}\left(1 - \frac{1}{R_{0i}}\right)\vec{e}_i.
\end{equation}
SIR:
\begin{equation}
\vec{I}^{*i}_u =\frac{\mu}{\beta_i}\left(R_{0i} - 1\right)\vec{e}_i.
\end{equation}
SIRS:
\begin{equation}
\vec{I}^{*i}_u = \frac{\mu\left(\mu + \gamma_i\right) S_0}{\delta_i\left(\mu + \gamma_i\right) - \gamma_i\eta_i}\left(1 - \frac{1}{R_{0i}}\right)\vec{e}_i.
\end{equation}
In contrast to the coupled models in which the condition $I^{*i} \geq 0$ required both $R_{0i} \geq 1$ and $R_{0i} > R_{0j}$ for all $j \triangleright i$, in all cases listed above a sufficient condition for the endemic solution $\vec{I}^{*i}_u$ to be non-negative is to have $R_{0i} \geq 1$. That is, the value of $R_{0i}$ relative to the remaining basic reproduction numbers $R_{0j}$ does not influence whether the endemic solution $\vec{I}_u^{*i}$ is physical.

In summary, the impact of introducing a coupling between the infectious compartments is to convert the next-generation matrix $\tens{K}$ from a strictly diagonal matrix (uncoupled case) to one which is lower triangular (coupled case). Although this does not change the expression for the basic reproduction number of each strain, nor the size of the susceptible populations at the infection-free and endemic equilibria, it does significantly convolute the endemic solutions $\vec{I}^{*i}$. Instead of being the eigenvectors of the diagonal matrix $\tens{K}_u$, which are just given by the canonical $n$-dimensional basis vectors $\vec{e}_i$, the endemic solutions in the coupled case become the eigenvectors of the generalized lower triangular matrix $\tens{B}^{-1}\tens{K}\tens{B}$ leading to endemic solutions with $I_j^{*i} > 0$ for all $j\triangleright i$.

\section{Conclusion}
\label{sec:conclusion}

In this article we have introduced a general framework to study coupled, multi-strain epidemic models designed to simulate the emergence and dissemination of mutational (e.g. drug-resistant) variations of pathogens. In particular, we have analyzed the properties of the infectious compartment network in these models and how they influence the system dynamics and solutions.


Firstly, we found that the introduction of a unidirectional coupling term (which we argue is the most general coupling available for a strictly \textit{time-forward} evolutionary process) results in the same generic functional form for $R_{0i}$ as that derived in canonical uncoupled models; it is simply the product of the mean per capita infection rate per unit time ($\beta_i S_0$) and the duration of the infectious period ($1/\delta_i$). This is an important result that shows that the reproductive capacity of each strain is independent of the mutation rates between them.

We then determined the necessary and sufficient conditions for epidemic fade-out and take-off, which we found can be stated in terms of the maximum basic reproduction number $R_{0f} \equiv \max_i\, R_{0i}$. As expected, the infected population of all strains decays rapidly and is extinguished asymptotically if $R_{0f} \leq 1$. Conversely, if $R_{0f} > 1$, the disease will spread and an epidemic occurs. Whilst these findings are similar to those found previously in uncoupled multi-strain epidemic models, we discussed how the coupling between the different strains in our model enlarges the space of initial conditions that lead to an epidemic outbreak: whereas in the former case the initial infectious population of the fittest strain $f$ must itself be positive, in the latter case any ancestor of strain $f$ can provide the initial spark to ignite the epidemic.


Next, by demonstrating that the system naturally evolves towards the equilibrium point that minimizes the susceptible population, we were able to show that when $R_{0f} > 1$, many strains with suboptimal reproductive capacity are driven to extinction --- these strains cannot be sustained by the limited pool of available susceptibles. Meanwhile, the fittest strain, along with each of its descendants, persist indefinitely. To support this finding, we also showed that only those endemic equilibria $P^{*i}$, for which $R_{0i} > 1$ \textit{and} $R_{0i} > R_{0j}$ for all $j$ that are descendants of  $i$, lie in the nonnegative orthant (i.e. are physically acceptable).


This, of course, is akin to the principle of competitive exclusion which asserts that several species competing over a shared resource cannot persist indefinitely: eventually, a single species rises to dominance at the exclusion (i.e. extinction) of the others. However, in the coupled model considered here, the persistence of strain $f$ ensures each of the strains descendant from $f$ survive indefinitely. In effect, as long as the fittest strain $f$ remains in circulation, it will continue to ``fuel'' each of its mutant descendants.

Lastly, we found that the fittest strain is not necessarily the most prevalent at equilibrium --- a result that superficially is not so surprising, given that the ancestor strain survives purely on direct transmission whilst its descendants are fueled by a combination of direct transmission \textit{and} amplification. However, we have shown that even in the absence of direct transmission of the descendant strain, amplification alone can cause it to become more prevalent than its fitter ancestor. This finding is particularly relevant for public health policy and planning since we have shown that a descendant strain can ``replace'' its ancestor (i.e. become more prevalent) even if it has a negligible basic reproduction number.

Combining these results we conclude that the mutation events that link the separate strains in the infectious compartment network are responsible for both the germination of new mutational variations, and (possibly) their sustained existence. However, the mutation rates themselves have no influence on the reproductive capacity of each strain and hence, the line of descent that comes to dominate (i.e. persist) at equilibrium. 

These findings can have important implications for selecting appropriate intervention or treatment strategies. In the context of antimicrobial drug resistance, the provision of treatment has been, and in some cases still remains, a controversial issue~\citep{Castillo-ChavezFeng1997}. As mentioned in the introduction, the proliferation of drug-resistant pathogen strains can often be traced back to the misuse and overuse of antimicrobials. Hence, it is natural to question whether increasing treatment rates further may actually produce adverse outcomes. Whilst that still remains a legitimate concern, because suboptimal treatment could produce new cases exhibiting existing patterns of drug resistance \textit{and} lead to the emergence of new, possibly more extensively drug-resistant strains, we have shown that the mutation rate itself does not affect the reproductive capacity of the mutant descendants. Therefore, on the one hand increasing treatment rates will not increase the potential of an epidemic outbreak of \textit{existing} strains, but on the other hand it may increase the likelihood of an even more prolific, i.e. fitter, strain emerging.

Finally, we must keep in mind that the conclusions outlined above have been made in a strictly deterministic setting. It would be interesting to investigate how stochastic effects may influence our findings particularly since 1. mutation events are an inherently random process, and 2. the initial numbers of individuals that acquire each mutant variant will presumably be small initially, thereby increasing the likelihood of strain extinction. Fortunately, the present framework can be readily adapted to incorporate stochastic events and this will serve as an important direction of future research.

In summary, this article has focused primarily on the mathematical properties and solutions of coupled, multi-strain epidemic models and as such should be viewed as a preliminary step towards understanding the genetic evolution of infectious diseases and the growing threat of antimicrobial drug resistance. In future works, we plan to utilize these results in a more practical context to fully explore the underlying mechanisms driving the emergence of drug-resistant pathogen strains of specific infectious diseases.



\bibliographystyle{elsarticle-num}  

\bibliography{references}   


\end{document}